\documentclass[preprint,12pt]{elsarticle}

\usepackage{amsmath, amssymb, amsfonts}
\usepackage{bm}
\usepackage{mathtools}
\usepackage{enumitem}
\usepackage{hyperref}

\newcommand{\E}{\mathbb{E}}
\newcommand{\Prob}{\mathbb{P}}

\journal{}

\makeatletter
\def\ps@pprintTitle{%
  \let\@oddhead\@empty
  \let\@evenhead\@empty
  \let\@oddfoot\@empty
  \let\@evenfoot\@oddfoot}
\makeatother

\begin{document}

\begin{frontmatter}

\title{Heuristic Solutions for the Best Secretary Problem}

\author{Eugene Seong}
\address{Department of Statistics, Korea University, Seoul, Republic of Korea}
\ead{eugenealices@gmail.com}

\begin{abstract}
This paper introduces a heuristic framework for the Best Secretary Problem, where one item must be selected using rank information only. We develop five data-responsive rules extending classical fixed-cutoff methods: an expected-record threshold, an adaptive deviation correction, a probabilistic early-accept rule, a two-phase relaxation, and a local dynamic programming approximation. These rules adjust thresholds sequentially as information accumulates. Simulations across diverse sample sizes, distributions, and autocorrelated settings show that the heuristics match or exceed traditional optimal rules in stability and efficiency. The expected-record rule remains strong despite its simplicity, the adaptive correction performs well under asymmetry, and the adaptive and probabilistic rules reduce average stopping times. An ensemble combining multiple rules yields the most stable performance. Overall, a few intuitive parameters achieve near-optimal results, demonstrating that data-responsive heuristics can effectively extend rank-based optimal stopping to dynamic decision environments.
\end{abstract}


\end{frontmatter}

% -------------------- Main Body --------------------

\section{Introduction}

The Best Secretary Problem (or \emph{Best-Choice Problem}) is a classical optimal stopping problem in which, using rank information only, one must select the maximum among a sequence with a single irrevocable choice. Given candidates (or prizes) \(X_1, X_2, \dots, X_n\) revealed sequentially, the decision maker at each time observes only past relative ranks and must either accept the current candidate immediately or reject it permanently. The goal is to choose a stopping time \(\tau\) that maximizes
\[
  \Prob\!\bigl(X_\tau = \max(X_1,\dots,X_n)\bigr).
\]

Gilbert and Mosteller (1966) showed that the strategy of observing approximately \(n/e\) candidates (about 37\%) and then accepting the first record (new maximum) thereafter is optimal. This “\(n/e\) rule” (the \emph{37\% rule}) has become a canonical expression of the exploration-exploitation balance.

A key feature of the problem is that the decision maker does not know the candidates’ values or their distribution; only rank information is observed. Hence monotone transformations of the underlying values do not affect the structure of optimal rules, and the problem is distributionally invariant in theory.

Classical solutions use a \emph{fixed cutoff} precomputed from the model; we refer to these as \emph{analytical cutoff rules}. The exact optimal rule of Gilbert \& Mosteller (1966) and the odds-sum rule of Bruss (2000) serve as theoretical baselines. Both compute a time-invariant cutoff \(r\) and accept the first record after \(r\).

In many practical settings, however, accumulated data call for thresholds that adapt to observed patterns. We term such approaches \emph{state-dependent heuristics} or \emph{adaptive rules}, where the cutoff responds to the current state (e.g., observed number of records or remaining horizon) or where early acceptance is allowed probabilistically. While related decision structures appear in behavioral decision making (Bearden, 2006), online learning (Lindley, 1961), and real-time control (Ferguson \& Bruss, 1993), their formulation in the rank-based secretary setting has been less systematic.

We introduce these ideas into the Best Secretary Problem and compare \emph{time-invariant} and \emph{data-responsive} thresholds within a common evaluation framework. The former provides theoretical optimality baselines; the latter targets adaptive robustness and practical efficiency. Although distributional invariance holds theoretically, real applications may involve dependence, shifting distributions, or online updates that affect stability. We therefore also assess empirical robustness under diverse distributions and autocorrelation.

The rules considered are as follows.  
In section 2 we review two conventional baselines:
\begin{itemize}[leftmargin=1.2em]
  \item \textbf{Exact optimal rule} — the \(n/e\)-based optimal cutoff of Gilbert \& Mosteller (1966).
  \item \textbf{Odds-sum rule} — Bruss (2000), stopping when the sum of odds reaches one.
\end{itemize}
In section 3 we present \emph{state-dependent heuristics}:
\begin{itemize}[leftmargin=1.2em]
  \item \textbf{Expected-record rule} — chooses a cutoff so that the expected number of records after it equals one.
  \item \textbf{Adaptive rule} — adjusts the cutoff in real time using the deviation between observed and expected records.
  \item \textbf{Probabilistic rule} — allows early acceptance in the exploration phase with increasing probability.
  \item \textbf{Two-phase rule} — introduces a relaxation phase between exploration and exploitation.
  \item \textbf{Rolling DP rule} — applies a local DP cutoff for short remaining horizons, otherwise uses a stable odds-based cutoff.
  \item \textbf{Voting ensemble} — combines the above rules via majority voting.
\end{itemize}

This contrast clarifies how time-invariant and data-responsive thresholds perform under the same rank-based observation model. Section 2 describes the analytical baselines; Section 3 introduces five heuristics and their ensemble; Section 4 reports simulation results across distributions and dependence; Section 5 concludes.

\section{Conventional Secretary Rules}

These rules share a single-cutoff structure that separates an observation phase from a selection phase: observe candidates up to a fixed time to set a reference record, then accept the first record thereafter. They serve as baselines in accuracy, generality, and simplicity against which our state-dependent heuristics in section 3 are compared.

\subsection{Exact Optimal Rule}
The \emph{Exact Optimal Rule} (Gilbert \& Mosteller, 1966) assumes candidates arrive in random order and only rank information is available. The optimal cutoff \(r^\ast\) maximizes
\begin{equation}
  P_r \;=\; \frac{r}{n}\sum_{k=r+1}^{n}\frac{1}{k-1}.
\end{equation}
The optimal strategy is to skip the first \(r^\ast\) candidates and then accept the first record thereafter. In summary:
\begin{itemize}[leftmargin=1.2em]
  \item \textbf{Observation phase:} Observe the first \(r^\ast\) candidates and set the maximum among them as the reference record.
  \item \textbf{Selection phase:} Accept the first subsequent candidate that exceeds the reference record.
\end{itemize}
As \(n\to\infty\), \(r^\ast/n \to 1/e \approx 0.3679\). The rule is often called the \(n/e\) or “37\%” rule and provides a closed-form benchmark.

\subsection{Odds-Sum Rule}
Bruss (2000) generalized secretary-type problems to stochastic time sequences, proposing the intuitive “sum the odds to one and stop” rule. For success probabilities \(p_i\) and odds \(o_i = p_i/(1-p_i)\), stop at the smallest \(s^\ast\) such that
\begin{equation}
  \sum_{i=s^\ast}^{n} o_i \;\ge\; 1.
\end{equation}
With equal \(p_i\), the resulting cutoff is close to the classic \(n/e\) rule; with heterogeneous \(p_i\), the odds-sum rule remains optimal in a wider class of sequences.

\section{Heuristic Rules}

We extend classical rules by introducing five data-responsive heuristics and one ensemble method that keep the exploration-exploitation logic but allow thresholds or acceptance conditions to adapt.

\subsection{Expected-Record Rule}
The expected-record rule reinterprets the cutoff via the \emph{expected number of records} after the cutoff (Ferguson, 1989). Define \(r\) such that the expected number of records after \(r\) equals one:
\begin{equation}
  \E[\text{records after } r] \;=\; H_n - H_r \;=\; \sum_{k=r+1}^{n} \frac{1}{k},
\end{equation}
where \(H_n = \sum_{k=1}^{n} 1/k\) is the \(n\)-th harmonic number. Setting
\begin{equation}
  H_n - H_r \;=\; 1
\end{equation}
yields a simple cutoff rule: after \(r\), accept the first record. The cutoff is easy to compute and, for large \(n\), lies near \(n/e\), giving performance close to the exact rule. While not theoretically optimal, its simplicity and efficiency make it a useful approximate baseline for the adaptive, probabilistic, and two-phase rules.
Although we list it among heuristic rules for consistency of exposition, the expected-record rule is essentially analytic rather than adaptive, as its cutoff is determined by a closed-form expectation rather than observed data.
We include it here to distinguish between the purely theoretical cutoff rules in Section 2 and the practically oriented heuristic and adaptive methods proposed in this study.

\subsection{Adaptive Deviation-Corrected Rule}
Instead of a fixed \(n/e\) cutoff, this rule adjusts the cutoff in real time using the deviation between the observed number of records and its expectation. Bearden (2006) argued that people adapt exploration length based on past outcomes, which can be formalized as an adaptive threshold.

Let \(r_0 = \lfloor n/e \rfloor\) and \(R\) be the number of records observed up to \(r_0\). With \(H_{r_0} = \sum_{k=1}^{r_0} 1/k\), define the adjusted cutoff
\begin{equation}
  r_{\mathrm{adj}} \;=\; r_0 - \gamma \,\bigl(R - H_{r_0}\bigr),
\end{equation}
with adjustment strength \(\gamma>0\), rounded to \(1 \le r_{\mathrm{adj}} \le n-1\). If \(R>H_{r_0}\), records arrive more often than expected (flatter distribution), so the cutoff is moved earlier; if \(R<H_{r_0}\), it is delayed.

Two modes are possible. In the \emph{batch mode}, compute \(r_{\mathrm{adj}}\) once at \(t=r_0\) and then use it as a fixed cutoff. In the \emph{streaming mode}, update at each time \(t\) using
\[
s_t \;=\; r_0 - \gamma\,(R_t - H_t),
\]
adapting the cutoff online. We use the streaming mode with numerical damping: clip the adjustment by \(\pm \text{cap}\,\sqrt{\log t}\) to limit early oscillations. In our experiments, \(\gamma=5\) and \(\text{cap}=3\).

\subsection{Probabilistic Early-Accept Rule}
Classical rules reject all candidates before the cutoff. Here we allow stochastic early acceptance during exploration (Lindley, 1961). Let \(\tau_0 = \lfloor n/e \rfloor\). If a record appears at time \(t<\tau_0\), accept it with probability
\begin{equation}
  q_t = \eta \left( \frac{t}{\tau_0} \right)^p, \qquad 0 < \eta < 1,
\end{equation}
where \(\eta\) controls the upper level and \(p>0\) controls curvature (linear when \(p=1\); back-loaded when \(p>1\)). We use \(\eta=0.05\) and \(p=2\). For \(t\ge\tau_0\), accept the first record deterministically. This relaxes a rigid cutoff and can improve exploration efficiency by occasionally accepting promising early records.

\subsection{Two-Phase Rule}
We introduce a three-stage structure--exploration, relaxation, and exploitation--by inserting a probabilistic relaxation phase between exploration and exploitation (cf. Bearden, 2006). Using the optimal cutoff \(r^\ast(n)\), define
\[
r^\ast = r^\ast(n), \qquad
r_1 = \lfloor 0.90\, r^\ast \rfloor, \qquad
r_2 = \min(n-1,\, r^\ast + \lfloor c_s \sqrt{n} \rfloor),
\]
with expansion factor \(c_s>0\). The acceptance probability at time \(t\) is
\[
q_t =
\begin{cases}
0, & t \le r_1,\\[6pt]
q_0 + (q_1 - q_0)\dfrac{t - r_1}{\max(1,\, r_2 - r_1)}, & r_1 < t \le r_2,\\[10pt]
1, & t > r_2,
\end{cases}
\]
so that Phase 1 is strict exploration, Phase 2 allows record acceptance with increasing probability, and Phase 3 accepts the first record deterministically. We use
\[
c_s = 0.4,\quad q_0 = 0.02,\quad q_1 = 0.30.
\]
This softens the transition at the cutoff and yields a smoother, more stable decision process in noisy environments.

\subsection{Rolling Local-DP Rule}
This rule approximates dynamic programming by decomposing the problem along the remaining horizon. With remaining length \(m = n - t + 1\), use the exact cutoff \(r^\ast(m)\) for small \(m \le m_0\) and a stable odds-based cutoff for \(m>m_0\):
\[
\begin{aligned}
\text{if } m > m_0: &\quad \text{use an odds-based first-record rule}, \\[3pt]
\text{else: } &\quad t_{\mathrm{local}} > r^\ast(m)\ \text{and record} \ \Rightarrow\ \text{accept}.
\end{aligned}
\]
Tabulate \(r^\ast(m)\) for \(m=1,\dots,m_0\); we use \(m_0=12\). This hybrid preserves exact logic on short horizons while keeping efficiency and stability for large \(n\).

\subsection{Voting Ensemble Rule}
Finally, we combine the seven rules (Exact, Odds, Expected-record, Adaptive, Probabilistic, Two-phase, Rolling-DP) by majority voting (Clemen, 1989). Let \(t_k\) denote the stopping time proposed by rule \(k\). Let \(r^{\mathrm{start}}_k\) be the earliest time at which rule \(k\) is willing to stop (i.e., begins making accept/reject decisions). Define the common start time
\[
  T_{\mathrm{common}} = \max_k \bigl(r^{\mathrm{start}}_k\bigr),
\]
and set
\begin{equation}
  t_{\mathrm{ensemble}} 
    = \min \Bigl\{\, t \ge T_{\mathrm{common}} 
       \;\Big|\; \#\{k : t_k = t\} \ge \lceil K/2 \rceil \Bigr\},
\end{equation}
the earliest time at which a strict majority recommends stopping. This leverages complementary strengths and increases stability under distributional uncertainty.

\section{Simulation}
\label{sec:simulation}

We compare eight rules: the analytical baselines (\emph{Exact}, \emph{Odds}); five state-dependent heuristics (\emph{Expected-record}: ER, \emph{Adaptive}: AD, \emph{Probabilistic}: PR, \emph{Two-phase}: TP, \emph{Rolling DP}: DP); and the \emph{Voting Ensemble} (VE).

\subsection{Simulation Design}
\label{subsec:design}

Sample sizes are $n \in \{50,100,200,500,1000\}$ with $B=10^4$ Monte Carlo replications. Distributions include symmetric \textsf{Uniform}$(0,1)$ and \textsf{Normal}$(0,1)$, and asymmetric \textsf{Exponential}$(1)$. To assess robustness beyond independence, we also consider \textsf{AR(1)} with $X_t=\phi X_{t-1}+\varepsilon_t$, $\varepsilon_t\sim\mathrm{N}(0,1)$, $\phi=0.5$. Because the problem is rank-based, monotone transformations and value scales do not affect the decision structure.

For each rule \(k\) we compute:
\begin{enumerate}[label=(\roman*), leftmargin=1.5em]
  \item \textbf{Success rate:}
  \[
    P_k = \frac{1}{B} \sum_{b=1}^{B} \mathbb{I}\!\bigl(X^{(b)}_{\tau_k^{(b)}} = \max_i X^{(b)}_i\bigr).
  \]
  \item \textbf{Average stopping time:}
  \[
    \bar{t}_k = \frac{1}{B} \sum_{b=1}^{B} \tau_k^{(b)}.
  \]
\end{enumerate}
Here \(\tau_k^{(b)}\) is the stopping time of rule \(k\) in replication \(b\). A trial counts as success if the chosen index equals the index of the true maximum.

\subsection{Results}

Tables~1--4 report \emph{success rates} by distribution. Bold entries mark the top three for each \(n\); bold underlined entries mark the top one. The last column shows the sum of ranks (smaller is better).

Overall, the Exact rule and the heuristics ER, AD, and the VE perform best. The ER rule is remarkably strong given its simplicity. The TP rule is generally weaker, notably under Normal and AR(1).

Across distributions, rules are close under Uniform and Normal. Under the asymmetric Exponential(1) distribution, the VE rule consistently outperforms the others, suggesting that the voting mechanism helps offset rule-specific biases. The AD rule also performs well under asymmetry. Under the AR(1) dependence, the overall performance of all rules declines because rank information becomes less informative, yet the ER rule remains relatively robust.

Tables~5--8 show \emph{average stopping times}. Bold entries mark the minimum and maximum in each column. Stopping times scale roughly linearly with \(n\), with modest differences across rules. AD and PR tend to stop earlier on average, while TP consistently stops latest, reflecting delays in its relaxation phase. Under the AR(1) dependence, variability in stopping times increases as early observations become less informative. Nevertheless, the AD rule remains fast and stable, demonstrating the robustness of its dynamic correction mechanism.

In summary, the Exact, ER, AD, and VE rules generally achieve the highest success rates. As expected, performance declines under dependence, yet the ER and VE rules remain comparatively robust.

\begin{table}[htbp]
\centering
\caption{Success Rate by Rule (\%) in Uniform(0,1)}
\begin{tabular}{lrrrrrr}
\hline
Rule & $n=50$ & 100 & 200 & 500 & 1000  & Sum(rank) \\
\hline
1) Exact & 37.12 & \textbf{37.32} & 36.63 & 36.46 & \textbf{37.21} & \textbf{22} \\
2) Odds & \textbf{37.33} & 37.20 & \textbf{\underline{37.41}} & 36.78 & \textbf{37.18} & \textbf{18} \\
3) ER & \textbf{\underline{37.47}} & \textbf{37.42} & \textbf{37.06} & 36.97 & 36.90 & \textbf{\underline{16}} \\
4) AD & 36.57 & 36.82 & 36.68 & 36.85 & \textbf{\underline{37.40}} & \textbf{22} \\
5) PR & 36.33 & \textbf{\underline{37.49}} & 36.25 & 37.10 & 36.77 & 25 \\
6) TP & 37.05 & 36.50 & 35.98 & \textbf{37.13} & 36.57 & 29 \\
7) DP & 36.54 & 36.31 & \textbf{37.19} & \textbf{\underline{37.23}} & 37.02 & \textbf{22} \\
8) VE & \textbf{37.44} & 36.78 & 35.95 & \textbf{37.16} & 36.56 & 26 \\
\hline
\end{tabular}
\end{table}

\begin{table}[htbp]
\centering
\caption{Success Rate by Rule (\%) in Normal(0,1)}
\begin{tabular}{lrrrrrr}
\hline
Rule & $n=50$ & 100 & 200 & 500 & 1000  & Sum(rank) \\
\hline
1) Exact & \textbf{\underline{37.89}} & \textbf{\underline{37.35}} & \textbf{\underline{37.09}} & 36.13 & \textbf{37.25} & \textbf{\underline{13}} \\
2) Odds & 36.70 & 36.60 & 36.76 & 36.46 & 36.90 & 23 \\
3) ER & \textbf{36.98} & 36.78 & \textbf{36.80} & 36.06 & 36.90 & 22 \\
4) AD & 36.54 & 36.41 & 36.70 & \textbf{36.58} & \textbf{\underline{37.53}} & \textbf{20} \\
5) PR & \textbf{37.23} & \textbf{37.01} & 36.30 & \textbf{36.54} & 36.31 & \textbf{20} \\
6) TP & 35.82 & \textbf{36.87} & 36.28 & 36.34 & 35.40 & 32 \\
7) DP & 35.91 & 36.07 & 36.19 & \textbf{\underline{36.69}} & 36.39 & 30 \\
8) VE & 36.53 & 36.63 & \textbf{36.88} & 36.51 & \textbf{37.51} & \textbf{19} \\
\hline
\end{tabular}
\end{table}

\begin{table}[htbp]
\centering
\caption{Success Rate by Rule (\%) in Exponential(1)}
\begin{tabular}{lrrrrrr}
\hline
Rule & $n=50$ & 100 & 200 & 500 & 1000  & Sum(rank) \\
\hline
1) Exact & \textbf{37.26} & 36.73 & \textbf{\underline{37.76}} & 36.66 & 36.74 & \textbf{20} \\
2) Odds & \textbf{37.74} & 36.23 & 37.07 & 35.97 & 36.20 & 29 \\
3) ER & 37.25 & \textbf{37.03} & \textbf{37.45} & 36.09 & \textbf{37.00} & \textbf{20} \\
4) AD & 36.62 & \textbf{\underline{37.69}} & 36.99 & \textbf{\underline{37.25}} & \textbf{37.34} & \textbf{17} \\
5) PR & 36.75 & \textbf{37.54} & 36.15 & 36.83 & 36.41 & 26 \\
6) TP & 37.17 & 36.75 & 37.05 & 36.13 & 36.35 & 27 \\
7) DP & 36.61 & 36.19 & 36.59 & \textbf{36.93} & \textbf{\underline{37.43}} & 26 \\
8) VE & \textbf{\underline{37.80}} & 36.75 & \textbf{37.67} & \textbf{36.93} & 36.91 & \textbf{\underline{13}} \\
\hline
\end{tabular}
\end{table}

\begin{table}[htbp]
\centering
\caption{Success Rate by Rule (\%) in AR(1)}
\begin{tabular}{lrrrrrr}
\hline
Rule & $n=50$ & 100 & 200 & 500 & 1000  & Sum(rank) \\
\hline
1) Exact & 29.76 & 31.23 & \textbf{32.89} & \textbf{34.39} & 34.36 & 22 \\
2) Odds & \textbf{30.71} & \textbf{32.22} & 32.16 & \textbf{\underline{34.68}} & 33.85 & \textbf{20} \\
3) ER & \textbf{30.44} & \textbf{\underline{32.45}} & \textbf{\underline{33.49}} & 33.26 & 34.18 & \textbf{\underline{17}} \\
4) AD & 29.87 & 31.17 & 32.31 & \textbf{34.47} & \textbf{34.48} & 22 \\
5) PR & 30.02 & 30.95 & 32.43 & 33.51 & \textbf{\underline{34.54}} & 23 \\
6) TP & 30.11 & 31.82 & 32.16 & 32.89 & 33.71 & 32 \\
7) DP & 29.81 & \textbf{32.23} & \textbf{32.86} & 33.87 & 34.12 & 23 \\
8) VE & \textbf{\underline{31.17}} & 30.94 & 32.41 & 34.28 & \textbf{34.43} & \textbf{21} \\
\hline
\end{tabular}
\end{table}

\begin{table}[htbp]
\centering
\caption{Average Stopping Time in Uniform(0,1)}
\begin{tabular}{lrrrrr}
\hline
Rule & $n=50$ & 100 & 200 & 500 & 1000 \\
\hline
1) Exact & 37.68 & 75.19 & 147.28 & 368.04 & 737.92 \\
2) Odds & 37.69 & 75.13 & 148.01 & 369.66 & 733.84 \\
3) ER & 37.63 & 73.87 & 147.21 & 367.12 & 739.93 \\
4) AD & \textbf{33.66} & \textbf{69.98} & 144.01 & 364.80 & 729.75 \\
5) PR & 35.26 & 70.55 & \textbf{143.67} & \textbf{361.30} & \textbf{719.85} \\
6) TP & \textbf{38.91} & \textbf{78.55} & \textbf{157.44} & \textbf{392.11} & \textbf{784.38} \\
7) DP & 37.65 & 74.92 & 147.98 & 367.52 & 734.69 \\
8) VE & 37.63 & 75.18 & 147.98 & 370.98 & 737.25 \\
\hline
\end{tabular}
\end{table}

\begin{table}[htbp]
\centering
\caption{Average Stopping Time in Normal(0,1)}
\begin{tabular}{lrrrrr}
\hline
Rule & $n=50$ & 100 & 200 & 500 & 1000 \\
\hline
1) Exact & 37.65 & 74.85 & 147.86 & 369.66 & 735.86 \\
2) Odds & 37.60 & 75.05 & 148.90 & 368.96 & 738.49 \\
3) ER & 37.76 & 74.19 & 147.98 & 370.03 & 736.66 \\
4) AD & \textbf{33.86} & \textbf{69.78} & 143.68 & 366.01 & 730.42 \\
5) PR & 34.99 & 70.08 & \textbf{143.61} & \textbf{359.51} & \textbf{720.30} \\
6) TP & \textbf{39.37} & \textbf{78.39} & \textbf{157.01} & \textbf{391.03} & \textbf{788.76} \\
7) DP & 37.83 & 75.47 & 148.29 & 368.47 & 737.45 \\
8) VE & 37.62 & 75.38 & 147.92 & 370.89 & 735.54 \\
\hline
\end{tabular}
\end{table}

\begin{table}[htbp]
\centering
\caption{Average Stopping Time in Exponential(1)}
\begin{tabular}{lrrrrr}
\hline
Rule & $n=50$ & 100 & 200 & 500 & 1000 \\
\hline
1) Exact & 37.79 & 75.11 & 147.24 & 368.53 & 739.29 \\
2) Odds & 37.60 & 74.91 & 147.84 & 367.07 & 740.75 \\
3) ER & 37.87 & 74.00 & 147.94 & 368.29 & 737.16 \\
4) AD & \textbf{33.87} & \textbf{69.85} & 144.35 & 363.22 & 730.90 \\
5) PR & 34.92 & 70.90 & \textbf{143.13} & \textbf{359.05} & \textbf{724.97} \\
6) TP & \textbf{39.16} & \textbf{78.78} & \textbf{156.98} & \textbf{393.96} & \textbf{785.64} \\
7) DP & 37.65 & 75.29 & 148.74 & 371.22 & 736.46 \\
8) VE & 37.60 & 75.18 & 148.33 & 369.69 & 735.05 \\
\hline
\end{tabular}
\end{table}

\begin{table}[htbp]
\centering
\caption{Average Stopping Time in AR(1)}
\begin{tabular}{lrrrrr}
\hline
Rule & $n=50$ & 100 & 200 & 500 & 1000 \\
\hline
1) Exact & 37.61 & 75.18 & 147.60 & 368.21 & 742.50 \\
2) Odds & 37.54 & 75.01 & 147.66 & 370.54 & 735.39 \\
3) ER & 37.55 & 74.08 & 147.95 & 368.70 & 738.60 \\
4) AD & \textbf{33.24} & \textbf{69.59} & \textbf{142.24} & 364.57 & 729.84 \\
5) PR & 34.87 & 71.14 & 143.24 & \textbf{359.55} & \textbf{725.07} \\
6) TP & \textbf{38.98} & \textbf{78.04} & \textbf{156.81} & \textbf{392.02} & \textbf{785.07} \\
7) DP & 37.71 & 74.76 & 148.30 & 367.72 & 736.61 \\
8) VE & 37.64 & 75.13 & 148.06 & 368.08 & 732.20 \\
\hline
\end{tabular}
\end{table}

\section{Conclusion and Limitations}

We compared conventional fixed-cutoff rules with several data-responsive heuristics that update thresholds based on observed information. Simulations show that the simple harmonic-based expected-record rule and the adaptive deviation correction are particularly strong in efficiency and stability, while the voting ensemble is most consistent under asymmetric distributions or dependence. These findings suggest that combining multiple decision principles can be more robust to real-world uncertainty than relying on any single analytical rule. The average stopping time analysis further indicates that real-time corrections enable faster, efficient decisions. In short, even simple heuristics can achieve performance comparable to classical rules when the exploration-exploitation balance is adjusted dynamically.

Several limitations remain. First, our analysis assumes that candidates arrive in random order and that only rank information is observable. In practice, however, real-world decisions may involve dependencies among candidates, observation delays, or partial information. Second, heuristic parameters (e.g., adjustment strength, early-accept rates, local horizon length) are set empirically; theoretical guidance or adaptive learning of these parameters is a direction for future work.
Third, our evaluation focuses solely on the probability of correct selection. Extending the analysis to other performance metrics--such as regret, expected rank, or cumulative reward--would provide a more comprehensive assessment.
Finally, while the proposed methods are computationally efficient and interpretable, further verification is needed for complex environments such as nonstationary time series or online learning. Despite these limitations, the results indicate that compact, transparent designs can deliver robust performance and offer a practical path to extend dynamic stopping in rank-based tasks.

% -------------------- References --------------------
\bibliographystyle{elsarticle-num}

\end{document}